\documentclass[aps,prd,superscriptaddress,showpacs,amsmath,amssymb]{revtex4}
\usepackage{graphicx,epsf}
\usepackage{amsfonts}
\usepackage{amssymb}
\usepackage{mathrsfs}
\usepackage[dvips]{color}
\definecolor{red}{rgb}{1,0,0}

\newcommand{\be}{\begin{eqnarray}}
\newcommand{\ee}{\end{eqnarray}}

\begin{document}

\begin{flushright}
DO-TH-09/07
\end{flushright}

\title{Neutrino-antineutrino oscillations as a possible solution for the LSND and MiniBooNE anomalies?}

\author{Sebastian Hollenberg}
\email{sebastian.hollenberg@uni-dortmund.de}
\author{Octavian Micu}
\email{octavian.micu@tu-dortmund.de}
\author{Heinrich P\"as}
\email{heinrich.paes@uni-dortmund.de}
\affiliation{Fakult\"at f\"ur Physik, Technische Universit\"at Dortmund, D-44221 Dortmund, Germany}

\begin{abstract}
We investigate resonance structures in CPT and Lorentz symmetry violating neutrino-antineutrino oscillations in a
two generation framework. We work with four non-zero CPT-violating parameters that
allow for resonant enhancements in neutrino-antineutrino oscillation phenomena in vacuo which are suitably described in terms of charge conjugation eigenstates of the system. We study the relation between the flavor, charge conjugation and mass eigenbasis of neutrino-antineutrino oscillations and examine the interplay between the available CPT-violating parameter space and possible resonance structures. Eventually we remark on the consequences of such scenarios for neutrino oscillation experiments, namely possible solutions for the LSND and MiniBooNE anomalies.
\end{abstract}
\maketitle

\section{Introduction}

The recently released data from the MiniBooNE collaboration \cite{AguilarArevalo:2008rc, AguilarArevalo:2009xn} reveal a resonance-like excess of events in the low-energy neutrino channel, but do not show a deviation from the expected oscillation pattern in the antineutrino channel. The LSND collaboration \cite{Aguilar:2001ty} on the other hand observes an excess in the antineutrino channel.
In order to understand these yet unexplained anomalies in terms of new physics, neutrino oscillation scenarios with altered dispersion relations have recently received attention \cite{Pas:2005rb, Hollenberg:2009bq, Nelson:2007yq}.
\par
For example, neutrino oscillations with altered dispersion relations can arise in extra-dimensional scenarios,
where active neutrinos are confined to a four dimensional brane, while sterile neutrinos take shortcuts in the extra dimensions.
\par
In such scenarios one typically encounters different problems in accommodating experimental data.
Firstly, active-sterile neutrino oscillations produce too many sterile neutrinos and thereby would create a deficiency in active neutrinos which
is not seen by MiniBooNE.
Secondly, the resonance energies for sterile neutrinos and antineutrinos in models with sterile neutrinos taking shortcuts in additional spacetime
dimensions are found to be the same. As a result gravitationally driven active-sterile neutrino oscillations are not capable of accommodating the latest MiniBooNE results \cite{AguilarArevalo:2009xn}. This calls for a mechanism capable of realizing different resonance energies for neutrinos and antineutrinos.
Thirdly, the width of the resonance in shortcut scenarios is typically too broad to provide a good fit to the experimental data \cite{Huber}.
Last but not least we understand that there is a hint that the MiniBooNE results for the resonance-like anomaly in the $\nu_{\mu}$ data actually look more like a
$\nu_{\mu} \to \bar{\nu}_e$ conversion than $\nu_{\mu} \to \nu_e$ events \cite{Stancu}.
\par
Problems with establishing resonances for different energies in active-sterile neutrino oscillation models for neutrinos and antineutrinos
as well as the problem of too broad resonance peaks can at least be partly rectified by incorporating matter effects in active-sterile neutrino oscillations with altered dispersion relations in the sterile neutrino sector \cite{Hollenberg:2009bq}. However, these
models can neither avoid the active neutrino deficiency due to $\nu_{a} \to \nu_s$ oscillations nor explain $\nu_{\mu} \to \bar{\nu}_e$ oscillation patterns.
\par
An alternative to active-sterile neutrino oscillations is a framework in which only active neutrinos contribute to oscillation phenomena and which allows for CPT-violating interactions. This avoids problems with active neutrino deficiencies and naturally permits $\nu_{\mu} \to \bar{\nu}_e$ events. Moreover it is conceivable that different resonance energies in neutrino-antineutrino oscillations are more easily established as compared to gravitationally determined active-sterile systems like
scenarios with sterile neutrino shortcuts.
\par
In section \ref{a model} and \ref{resonances} of this letter we study for the first time a two generation model for
neutrino-antineutrino oscillations assuming a minimal set of neutrino-antineutrino couplings. We show that this scenario gives rise to resonance structures in neutrino-antineutrino oscillations which are suitably described
in terms of charge conjugation $\mathscr{C}$ eigenstates.
As long as matter effects are not taken into account, and the coupling matrices are real, $\mathscr{C}$-even eigenstates do not mix with $\mathscr{C}$-odd eigenstates, but only with states of their own kind and vice versa.
$\mathscr{C}$-even and $\mathscr{C}$-odd eigenstates decouple in our model and we calculate effective mixing angles and eigenvalues of the Hamiltonian for the two systems with definite $\mathscr{C}$-parity. The two independent systems reveal resonant mixing for certain regions of the available parameter space. We find that it is possible to have $\mathscr{C}$-even and $\mathscr{C}$-odd resonances at the same time with different resonance energies. In addition we pin down the mixing between eigenstates of defined $\mathscr{C}$-parity and the mass eigenstates. In sections \ref{discussion} and \ref{discussion2} we examine the relation between common flavor and mass eigenstates and comment on the direction dependence of the four  CPT- and Lorentz-violating coefficients in our model. Eventually we remark on the consequences of our approach for the LSND and MiniBooNE anomalies.

\section{A Model for two generation neutrino-antineutrino mixing}\label{a model}

It has been shown by Kostelecky and Mewes \cite{Kostelecky:2003cr} that a minimal extension of Standard Model physics
including all possible CPT and Lorentz symmetry violating Dirac- and Majorana-type couplings
to left- and right-handed neutrinos gives rise to an effective Hamiltonian $h_{\text{eff}}$
describing neutrino-antineutrino oscillations. An explicit form for $h_{\text{eff}}$ can be
obtained under the assumption that low-energy physics is dominated by renormalizable Lorentz-violating
operators. In the case of two flavors the Lorentz-violating equations of motion can be written in
analogy to the Dirac equation
    \be
        (i\Gamma^{\nu}_{AB}\partial_{\nu} - M_{AB})\nu_B = 0,
    \ee
where $A, B = e,~\mu,~e^c,~\mu^c$ are flavor indices for neutrinos and antineutrinos respectively.
In this notation one may expand the coefficients $\Gamma^{\nu}_{AB}$ and $M_{AB}$ in terms of the
$\gamma$-matrices determining the Dirac algebra
    \be
        \Gamma^{\nu}_{AB} &=& \gamma^{\nu} \delta_{AB} + c^{\mu\nu}_{AB}\gamma_{\mu} + d^{\mu\nu}_{AB}
        \gamma_{5}\gamma_{\mu} + e^{\nu}_{AB} + if^{\nu}_{AB}\gamma_{5} + \frac{1}{2}g^{\lambda\mu\nu}_{AB}
        \sigma_{\lambda\mu}, \\
        M_{AB} &=& m_{AB} + im_{5AB}\gamma_{5} + a^{\mu}_{AB}\gamma_{\mu} + b^{\mu}_{AB}\gamma_{5}\gamma_{\mu}
        + \frac{1}{2}H^{\mu\nu}_{AB}\sigma_{\mu\nu}.
    \ee
The masses $m$ and $m_5$ are CPT and Lorentz conserving and equivalent to the standard Majorana- and Dirac-type
masses. The coefficients $c,~d,~H$ are CPT conserving, but Lorentz-violating; $a,~b,~e,~f,~g$ are both CPT- and Lorentz-violating.
It has been shown that this ansatz leads to an effective Hamiltonian in flavor space entailing a modification of the standard oscillation phenomena in neutrinos due to CPT- and Lorentz-violating
effects. For further details of the formalism the reader is referred to \cite{Kostelecky:2003cr, Choubey:2003ke}.
\par
Restricting ourselves to the first two neutrino generations, the propagation of kinematically allowed states $\nu_{\text{L}}^{a}$, $(\nu_{\text{L}}^{a})^{c}$ is governed by the Schr\"odinger equation
    \be
        i\frac{d}{dt}\left(\begin{array}{c} \nu_a \\ \nu_a^c\end{array}\right)
        = \sum\limits_{b=e, \mu}^{}(h_{\text{eff}})_{ab}
        \left(\begin{array}{c}\nu_b \\ \nu_b^c\end{array}\right).
    \ee
The subscript L indicating the chirality of the neutrinos will henceforth be omitted. Here $\nu_{i}$ are
2-spinors with the index $i=e,~\mu$ denoting the neutrino flavor; $c$ refers to charge conjugate neutrino states. In a two generation analysis
the Schr\"odinger equation assumes a form
    \be
        i\frac{d}{dt} \left(\begin{array}{c} \nu_e \\ \nu_{\mu} \\ \nu_e^c \\ \nu_{\mu}^c \end{array}\right)
        = \begin{pmatrix} \mathfrak{A}_{ee} & \mathfrak{A}_{e\mu} & \mathfrak{B}_{ee} & \mathfrak{B}_{e\mu} \\
                        \mathfrak{A}_{\mu e} & \mathfrak{A}_{\mu\mu} & \mathfrak{B}_{\mu e} & \mathfrak{B}_{\mu\mu} \\
                        \mathfrak{C}_{ee} & \mathfrak{C}_{e\mu} & \mathfrak{D}_{ee} & \mathfrak{D}_{e\mu} \\
                        \mathfrak{C}_{\mu e} & \mathfrak{C}_{\mu\mu} & \mathfrak{D}_{\mu e} & \mathfrak{D}_{\mu\mu} \\
                        \end{pmatrix}
        \left(\begin{array}{c} \nu_e \\ \nu_{\mu} \\ \nu_e^c \\ \nu_{\mu}^c \end{array}\right)
        \label{SGL}  ,
    \ee
where according to \cite{Kostelecky:2003cr} we specify
    \be
        \mathfrak{A}_{ab} &=& E \delta_{ab} + \frac{1}{2E} (m_l^{} m_l^{\dagger})_{ab} + \frac{1}{E} \left[
                 \left(a_{\text{L}}\right)^{\mu}p_{\mu} - \left(c_{\text{L}}\right)^{\mu\nu}p_{\mu}
                 p_{\nu}\right]_{ab}, \label{matrixA}\\
        \mathfrak{B}_{ab} &=& -\frac{i}{E}\sqrt{2}p_{\mu}\left(\epsilon_{+}\right)_{\nu} \left[\left(g^{\mu\nu\sigma}p_{\sigma}
                 - H^{\mu\nu}\right)\mathcal{C}\right]_{ab}, \label{matrixB}\\
        \mathfrak{C}_{ab} &=&\frac{i}{E}\sqrt{2}p_{\mu}\left(\epsilon_{+}\right)_{\nu}^{\ast}
                  \left[\left(g^{\mu\nu\sigma}p_{\sigma} + H^{\mu\nu}\right)\mathcal{C}\right]^{\ast}_{ab}, \label{matrixC}\\
        \mathfrak{D}_{ab} &=& E\delta_{ab} + \frac{1}{2E} (m_l^{} m_l^{\dagger})_{ab}^{\ast} + \frac{1}{E} \left[
                 -\left(a_{\text{L}}\right)^{\mu}p_{\mu} - \left(c_{\text{L}}\right)^{\mu\nu}p_{\mu}
                 p_{\nu}\right]^{\ast}_{ab} \label{matrixD},
    \ee
with $\left(a_{\text{L}}\right)^{\mu}_{ab} \equiv \left(a+b\right)^{\mu}_{ab}$, $\left(c_{\text{L}}\right)^{\mu\nu}_{ab}
\equiv \left(c+d\right)^{\mu\nu}_{ab}$ and $\mathcal{C}$ being the charge conjugation matrix in the chosen basis
with $\nu_A^c = \mathcal{C}_{AB}\nu_B$.
\par
In the following considerations we restrict our attention to a set of four non-zero coefficients. This set
allows for minimal neutrino-antineutrino oscillations, but also induces altered dispersion relations in the
standard $\nu_e$-$\nu_{\mu}$ and $\nu_e^c$-$\nu_{\mu}^c$ sector respectively:
\par\noindent
\newline
The usual Lorentz-conserving mass term in Eqs.~(\ref{matrixA}-\ref{matrixD}) may be parameterized in analogy
to two flavor Lorentz-conserving neutrino oscillations in vacuo
    \be
        (m_l^{} m_l^{\dagger})_{ab} = (m_l^{} m_l^{\dagger})_{ab}^{\ast} =
        \frac{\Sigma m^2}{2} + \frac{\Delta m^2}{2} \begin{pmatrix} -\cos2\theta & \sin2\theta \\
        \sin2\theta & \cos2\theta \end{pmatrix},
    \ee
where $\Sigma m^2 = m_1^2 + m_2^2$ and $\Delta m^2 = m_2^2 - m_1^2$ are given by the standard values which for the purpose of our work we take to be the solar ones.
The angle $\theta$ is the vacuum mixing angle which equals the solar mixing angle, if one only considers the first two neutrino generations.
\par
It can be shown that the only physically significant combinations involving $g$-type coefficients
are given by
    \be
        p_{\mu}(\epsilon_{+})_{\nu}g^{\mu\nu\sigma} = E(\epsilon_{+})_{\nu}\tilde{g}^{\nu\sigma},
    \ee
where $\tilde{g}^{\nu\sigma}$ is defined via
    \be
        \tilde{g}^{\nu\sigma} \equiv g^{0\nu\sigma} + \frac{i}{2}\epsilon^{0\nu}_{~~\gamma\rho}g^{\gamma\rho\sigma}.
    \ee
We further assume for definiteness and as an explicit example that
    \be
        \tilde{g}^{ZT} \neq 0 \label{gchoice}
    \ee
is the only non-zero $\tilde{g}$-type coefficient.
\par
This assumption is convenient to establish an easily workable model which already entails
interesting phenomenological consequences. At the end of section \ref{a model} we will comment on the implications of
possible extensions of our model.
\par
We choose a sun-centered celestial equatorial frame with coordinates $(T,X,Y,Z)$. In this frame the $Z$ direction is lying along
the earth's rotational axis. We choose a parametrization
    \be
        \hat{\vec{p}} &=& \left(\sin\Theta \cos\Phi,~\sin\Theta \sin\Phi,~\cos\Theta\right), \\
        \hat{\epsilon}_1 &=& \left(\cos\Theta \cos\Phi,~\cos\Theta \sin\Phi,~-\sin\Theta\right), \\
        \hat{\epsilon}_2 &=& \left(-\sin\Phi,~\cos\Phi,~0\right),
    \ee
where $\hat{\vec{p}}$ is the unit 3-vector of the particle's momentum, and find that a choice
    \be
        \left(\epsilon_+\right)^{\nu} =
        \frac{1}{\sqrt{2}}\left(0, \hat{\epsilon}_1 +i\hat{\epsilon}_2\right)
    \ee
analogous to the photon helicity basis is suitable in this context. Here $\Theta$ is the celestial colatitude and
$\Phi$ denotes the celestial longitude.
\par
We now restrict ourselves to a specific choice of parameters which leads to a particularly interesting scenario for phenomenological studies.
We define
    \be
        \tilde{g}^{ZT}_{ee^c} \equiv \tilde{g}^{ZT}_{e^ce} &=& ig_e, \label{gcoeff1}\\
        \tilde{g}^{ZT}_{\mu\mu^c} \equiv \tilde{g}^{ZT}_{\mu^c\mu} &=& ig_{\mu} \label{gcoeff2},
    \ee
where the $g_a$'s are real, and introduce $B_e$/$b_e$- and $B_{\mu}$/$b_{\mu}$-type coefficients as follows
    \be
        B_e(E) = 2\sin\Theta~g_e E^2 &\equiv& b_{e}E^2, \\
        B_{\mu}(E) = 2\sin\Theta~g_{\mu} E^2 &\equiv& b_{\mu}E^2.
    \ee
\par
Since $p_{\mu}$ in our model can approximately be identified with
$p_{\mu} = (E,-\vec{p}) \simeq E (1, -\hat{\vec{p}}) = E\hat{p}_{\mu}$, where $\hat{\vec{p}}$, $\hat{p}_{\mu}$
denote unit vectors, we rewrite the $c_{\text{L}}$-type coefficients via
    \be
        (c_{\text{L}})_{ab}^{\mu\nu}p_{\mu}p_{\nu} = E^2
        (c_{\text{L}})_{ab}^{\mu\nu}\hat{p}_{\mu}\hat{p}_{\nu} \equiv \frac{1}{2}c_{ab}E^2,
    \ee
where the trace $\eta_{\mu\nu}(c_{\text{L}})^{\mu\nu}_{ab}$ is unobservable due to the fact that it may
be absorbed into the usual kinetic term of the Hamiltonian. Setting the trace equal to zero results in
    \be
        c_{ab} = 2 (c_{\text{L}})_{ab}^{TT}[1+\cos^2\Theta],
    \ee
if we assume that $(c_\text{L})^{TT}_{ab}$ and $(c_\text{L})^{ZZ}_{ab}$ are the only non-zero $c_{\text{L}}$-type
coefficients. Those two coefficients are actually equal to one another due to the fact that they are assumed to be the only contributing matrix entries in a matrix with vanishing trace. The superscript $TT$ and subscript L for the $c_{\text{L}}$-type coefficients will henceforth be omitted.
We define $C_{ab}$/$c_{ab}$-type coefficients
    \be
        C_{e\mu}(E) \equiv C_{\mu e}(E) &=& 0, \\
        C_{ee}(E) &=& c_{ee}E^2, \\
        C_{\mu\mu}(E) &=& c_{\mu\mu} E^2,
    \ee
assuming that all $c_{ab}$'s are real.
\par\noindent
Moreover we neglect all contributions stemming from $a_{\text{L}}$- and $H$-type coefficients for the moment.
At this stage of the analysis we remark that both $B$- and $C$-type coefficients are direction dependent;
we will discuss this feature in section \ref{discussion}.
\par\noindent
\newline
The effective Hamiltonian now reads

    \be
        h_{\text{eff}} = &&\text{diag}\left(E+\frac{\Sigma m^2}{4E}\right) + \nonumber \\
        &+&\begin{pmatrix} -\frac{\Delta m^2}{4E}\cos2\theta - \frac{C_{ee}(E)}{2E} &
                          \frac{\Delta m^2}{4E}\sin2\theta  &
                          \frac{B_{e}(E)}{2E} & 0 \\
                          \frac{\Delta m^2}{4E}\sin2\theta  &
                          \frac{\Delta m^2}{4E}\cos2\theta - \frac{C_{\mu\mu}(E)}{2E} &
                          0 & \frac{B_{\mu}(E)}{2E} \\
                          \frac{B_{e}(E)}{2E} & 0 &
                          -\frac{\Delta m^2}{4E}\cos2\theta - \frac{C_{ee}(E)}{2E} &
                          \frac{\Delta m^2}{4E}\sin2\theta  \\
                          0 & \frac{B_{\mu}(E)}{2E} &
                          \frac{\Delta m^2}{4E}\sin2\theta  &
                          \frac{\Delta m^2}{4E}\cos2\theta - \frac{C_{\mu\mu}(E)}{2E}\\
           \end{pmatrix} \label{heff}.
    \ee
\newline

This matrix is symmetric and we may therefore interpret the eigenvalues of the given Hamiltonian as the
effective energy eigenvalues of the system. The effective Hamiltonian Eq.~(\ref{heff}) is a $4 \times 4$
matrix, which consists of four $2 \times 2$ block matrices. As can be easily seen, the CPT-violating $B$-coefficients generate neutrino-antineutrino mixing, while the CPT-conserving $C$-coefficients yield Lorentz-violating terms altering the dispersion relations of the flavor  states. In the limit where the CPT- and Lorentz-violating coefficients are set to zero the Hamiltonian which describes standard neutrino oscillations in vacuo is retained.
The explicit form of Eq.~(\ref{heff}) suggests
that it might be possible to block-diagonalize the effective Hamiltonian. We do indeed find a unitary
matrix
    \be
        U = \frac{1}{\sqrt{2}} \begin{pmatrix} 1 & 0 & 1 & 0 \\ 0 & 1 & 0 &1 \\ -1 & 0 & 1 & 0 \\
            0 & -1 & 0 & 1 \end{pmatrix} \label{U},
    \ee
which permits to express the effective Hamiltonian $h_{\text{eff}}$ in terms of a block-diagonal
effective Hamiltonian $\tilde{h}_{\text{eff}}$ via
    \be
        h_{\text{eff}} = U~\tilde{h}_{\text{eff}}~U^{\dagger}.
    \ee
If we plug this expression back into Eq.~(\ref{SGL}) we realize that it is convenient to perform a
change of basis according to
    \be
        \left(\begin{array}{c} \nu_e \\ \nu_{\mu} \\ \nu_e^c \\ \nu_{\mu}^c \end{array}\right) \to
        \frac{1}{\sqrt{2}} \begin{pmatrix} 1 & 0 & -1 & 0 \\ 0 & 1 & 0 & -1 \\ 1 & 0 & 1 & 0 \\
        0 & 1 & 0 & 1 \end{pmatrix}
        \left(\begin{array}{c} \nu_e \\ \nu_{\mu} \\ \nu_e^c \\ \nu_{\mu}^c \end{array}\right)
        = \frac{1}{\sqrt{2}} \left(\begin{array}{c} \nu_e^{} - \nu_e^c \\ \nu_{\mu}^{} - \nu_{\mu}^c \\
        \nu_e^{} + \nu_e^c \\ \nu_{\mu}^{} + \nu_{\mu}^c \end{array}\right).
    \ee
We define
    \be
        \nu^- &=& \left(\begin{array}{c} \nu_{e}^- \\ \nu_{\mu}^- \end{array}\right) \equiv \frac{1}{\sqrt{2}} \left(\begin{array}{c} \nu_e^{} - \nu_e^c \\ \nu_{\mu}^{} - \nu_{\mu}^c
        \end{array}\right) , \\
        \nu^+ &=& \left(\begin{array}{c} \nu_{e}^+ \\ \nu_{\mu}^+ \end{array}\right) \equiv \frac{1}{\sqrt{2}} \left(\begin{array}{c} \nu_e^{} + \nu_e^c \\ \nu_{\mu}^{} + \nu_{\mu}^c
        \end{array} \right) ;
    \ee
and introduce the charge conjugation operator $\mathscr{C}$ in flavor space which acts on the different
neutrino states as follows
    \be
        \mathscr{C} \nu_a^{} &=& \nu_a^c, \\
        \mathscr{C} \nu_a^c &=& \nu_a^{}.
    \ee
The aforementioned states $\nu^-$ and $\nu^+$ are eigenstates of the charge conjugation
operator in a way that
    \be
        \mathscr{C} \nu^- &=& -\nu^-, \\
        \mathscr{C} \nu^+ &=& + \nu^+
    \ee
holds. We will therefore refer to $\nu^-$ as a $\mathscr{C}$-odd state and to $\nu^+$ as a $\mathscr{C}$-even
state respectively.
\par
Given these prerequisites the Schr\"odinger equation for our effective Hamiltonian can be rewritten in a
$2 \times 2$ block matrix form
    \be
        i\frac{d}{dt} \left(\begin{array}{c} \nu^- \\ \nu^+ \end{array}\right) = \begin{pmatrix}
        h_{\text{eff}}^{\mathscr{C}\text{-odd}} & 0 \\ 0 & h_{\text{eff}}^{\mathscr{C}\text{-even}}
        \end{pmatrix} \left(\begin{array}{c} \nu^- \\ \nu^+ \end{array}\right) \label{SGLC}.
    \ee
It is obvious that neutrino-antineutrino mixing in our model is possible only between $\mathscr{C}$-even
states or $\mathscr{C}$-odd states. Since the two non-zero blocks in Eq.~(\ref{SGLC}) for $\mathscr{C}$-even and $\mathscr{C}$-odd states
do not mix the different neutrino-antineutrino superpositions
with definite $\mathscr{C}$-parity decouple and in order to study the characteristics of neutrino oscillations we analyze the blocks
$h_{\text{eff}}^{\mathscr{C}\text{-odd}}$ and $h_{\text{eff}}^{\mathscr{C}\text{-even}}$ separately. Moreover the $\mathscr{C}$-eigenstates
assume the role of flavor eigenstates in common neutrino oscillation phenomena. We will therefore refer to these "new" or "effective" flavor states $\nu^{\pm}$ as $\mathscr{C}$-flavor states.
\par\noindent
\newline
At this stage of our analysis we comment on the limitations of our approach:
\par
Firstly, the general formalism to describe CPT- and Lorentz-violating neutrino oscillations
Eqs.~(\ref{matrixA}-\ref{matrixD}) contains the $V-A$ coupling for
coherent elastic forward scattering of neutrinos in matter encoded in the $a_{\text{L}}$-type
coefficients. For illustrative purposes we restrict our attention to interactions between electrons in
matter and electron neutrinos.
The $e-e$ and $e^c-e^c$ element of our Hamiltonian therefore receives an extra contribution of
    \be
        \left(a_{\text{L}}\right)^{0}_{ee} &=& \sqrt{2} G_{\text{F}} n_e, \\
        \left(a_{\text{L}}\right)^{0}_{e^ce^c} &=& -\sqrt{2} G_{\text{F}} n_e,
    \ee
where $n_e$ is the electron number density in matter. We now apply the same unitary transformation
$U$ to the effective Hamiltonian containing matter potentials and find
    \be
        i\frac{d}{dt} \left(\begin{array}{c} \nu^- \\ \nu^+ \end{array}\right) = \begin{pmatrix}
        h_{\text{eff}}^{\mathscr{C}\text{-odd}} & M \\ M & h_{\text{eff}}^{\mathscr{C}\text{-even}}
        \end{pmatrix} \left(\begin{array}{c} \nu^- \\ \nu^+ \end{array}\right).
    \ee
In a basis of $\mathscr{C}$-eigenstates matter effects only contribute to the off-diagonal blocks of
$h_{\text{eff}}$. The deviation from the block-diagonal effective Hamiltonian is given by
    \be
        M = \begin{pmatrix} \sqrt{2} G_{\text{F}} n_e & 0 \\ 0 & 0\end{pmatrix}.
    \ee
We find that, in the given framework, standard matter effects introduce a mixing between
$\mathscr{C}$-even and $\mathscr{C}$-odd states in the electron sector. In turn this means that matter effects
invalidate an approach via $\mathscr{C}$-eigenstates and the diagonalization procedure becomes more involved.
\par\noindent
\newline
Secondly, it is possible to relax the choice Eq.~(\ref{gchoice}) in a way that we might choose different components of
$\tilde{g}^{\nu\sigma}$ to be non-zero. Taking into account the analogy of the $\left(\epsilon_+\right)^{\nu}$ vector to the
photon helicity basis we notice that this choice in general will give rise to complex-valued entries in the off-diagonal blocks in $\tilde{h}_{\text{eff}}$. In this case the overall factor of $i$ in the $\mathfrak{B}_{ab}$ and $\mathfrak{C}_{ab}$ coefficients
does not cancel any more. The $B_a$-type coefficients are now complex-valued.
Let us for illustrative purposes rewrite the effective Hamiltonian  in the following way
    \be
        h_{\text{eff}} = \begin{pmatrix} \mathfrak{A} & \mathfrak{B} \\ \mathfrak{B}^{\ast} &
        \mathfrak{A} \end{pmatrix},
    \ee
where $\mathfrak{A}$, $\mathfrak{B}$ are $2 \times 2$ matrices given in Eq.~(\ref{heff}) with the modified $B$-type coefficients.
The asterisk indicates complex conjugation.
We can now apply the unitary transformation matrix $U$ given in Eq.~(\ref{U}) and find
    \be
        \tilde{h}_{\text{eff}} = \begin{pmatrix} \mathfrak{A} - \text{Re}\mathfrak{B} & i~\text{Im}\mathfrak{B}
        \\ -i~\text{Im}\mathfrak{B} & \mathfrak{A} + \text{Re}\mathfrak{B} \end{pmatrix}.
    \ee
$\text{Re}$ and $\text{Im}$ here mean real and imaginary part and are understood as
    \be
        \text{Re}\mathfrak{B} &\equiv& \frac{1}{2} \left(\mathfrak{B} + \mathfrak{B}^{\ast}\right), \\
        \text{Im}\mathfrak{B} &\equiv& \frac{1}{2i} \left(\mathfrak{B} - \mathfrak{B}^{\ast}\right).
    \ee
In conclusion this means that not only the presence of standard matter effects introduces a coupling between
$\mathscr{C}$-even and $\mathscr{C}$-odd neutrino states, but also the complex phase in the neutrino-antineutrino
couplings. Choosing the matrix $\mathfrak{B}$ to be real then reproduces the choice guiding our subsequent analysis.

\section{Resonant two-state oscillations with definite $\mathscr{C}$-parity}\label{resonances}

Let us first of all consider oscillations between $\mathscr{C}$-odd states only. The effective $2 \times 2$
$\mathscr{C}$-odd Hamiltonian reads
    \be
        h_{\text{eff}}^{\mathscr{C}\text{-odd}} = \text{diag}\left(E+\frac{\Sigma m^2}{4E}\right)+
        \begin{pmatrix} -\frac{\Delta m^2}{4E}\cos2\theta -
        \frac{(b_e + c_{ee})E}{2} & \frac{\Delta m^2}{4E}\sin2\theta  \\
        \frac{\Delta m^2}{4E}\sin2\theta  & \frac{\Delta m^2}{4E}\cos2\theta -
        \frac{(b_{\mu} + c_{\mu \mu})E}{2}
        \end{pmatrix}
    \ee
and it is straightforward to diagonalize this system and for these purposes calculate an effective mixing angle
$\theta_{\mathscr{C}\text{-odd}}$ to give
    \be
        \tan2\theta_{\mathscr{C}\text{-odd}} = \frac{\Delta m^2 \sin2\theta}{(b_e-b_{\mu}+
        c_{ee}-c_{\mu\mu})E^2+\Delta m^2 \cos2\theta} \label{thetaodd}.
    \ee
\begin{figure}
\centering
\raisebox{6.5cm}{$\sin^22\theta_{\text{eff}}$}
\includegraphics[scale=1.3]{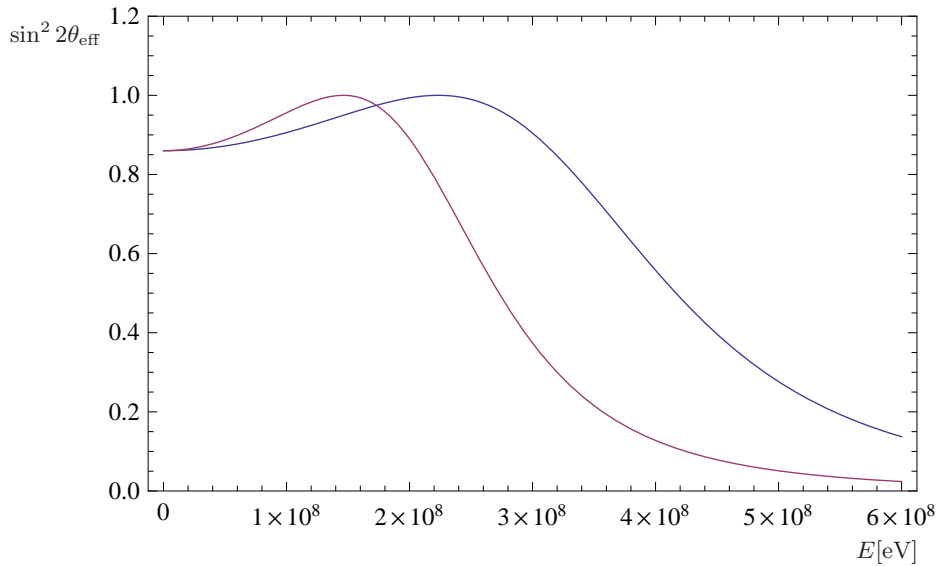}
\\{\hspace{11cm}$E[\text{eV}]$}
\caption{Resonance structures between charge conjugation eigenstates. Shown is the sine-squared of the effective mixing angles $\theta_{\mathscr{C}\text{-odd}}$ (blue curve) and
$\theta_{\mathscr{C}\text{-even}}$ (red curve). We choose $b_e=1\times 10^{-21}$, $b_{\mu}=0.6\times 10^{-21}$, $c_{\mu\mu}=3\times 10^{-21}$, $c_{ee}=2\times 10^{-21}$ for illustrative purposes. We take $\Delta m^2 = 8 \times 10^{-5} \text{eV}^2$ as well as $\sin^22\theta = 0.86$. In this case the resonance energy for the $\mathscr{C}\text{-odd}$ mixing is higher as compared to the resonance energy of the $\mathscr{C}\text{-even}$ mixing.} \label{sin1}
\end{figure}
One readily infers from Eq.~(\ref{thetaodd}) that the mixing between $\mathscr{C}$-odd states becomes maximal
for a resonance energy
    \be
        E^{\mathscr{C}\text{-odd}}_{\text{res}} = \sqrt{\frac{\Delta m^2 \cos2\theta}{b_{\mu}-b_e +
        c_{\mu\mu}- c_{ee}}} \label{Eresodd}.
    \ee
This resonance only exists if the condition $b_{\mu}-b_e + c_{\mu\mu}- c_{ee} > 0$ is satisfied.
The effective mass eigenvalues of the Hamiltonian are given by
    \be
        m_{1}^2 &=& \frac{\Sigma m^2}{2}-\frac{1}{2}[(b_e+b_{\mu}+c_{ee}+c_{\mu\mu})E^2 +\kappa_{\mathscr{C}\text{-odd}}], \label{mass1}\\
        m_{2}^2 &=& \frac{\Sigma m^2}{2}-\frac{1}{2}[(b_e+b_{\mu}+c_{ee}+c_{\mu\mu})E^2 -\kappa_{\mathscr{C}\text{-odd}}] \label{mass2},
    \ee
where
    \be
        \kappa_{\mathscr{C}\text{-odd}}^2 = (b_e - b_{\mu}+c_{ee}-c_{\mu\mu})^2 E^4 +
        2\Delta m^2 (b_e - b_{\mu}+c_{ee}-c_{\mu\mu})\cos2\theta E^2 + (\Delta m^2)^2 \label{kappaodd}.
    \ee
The diagonalization of $h_{\text{eff}}^{\mathscr{C}\text{-odd}}$ in the case under consideration connects
$\mathscr{C}$-eigenstates with mass eigenstates
    \be
        \left(\begin{array}{c} \nu_e^- \\ \nu_{\mu}^- \end{array}\right) = \begin{pmatrix}
        \cos\theta_{\mathscr{C}\text{-odd}} & \sin\theta_{\mathscr{C}\text{-odd}} \\
        -\sin\theta_{\mathscr{C}\text{-odd}} & \cos\theta_{\mathscr{C}\text{-odd}} \end{pmatrix}
        \left(\begin{array}{c} \nu_1 \\ \nu_2 \end{array}\right).
    \ee
The analysis of the $2 \times 2$ $\mathscr{C}$-even Hamiltonian
    \be
        h_{\text{eff}}^{\mathscr{C}\text{-even}} = \text{diag}\left(E+\frac{\Sigma m^2}{4E}\right)+
        \begin{pmatrix} -\frac{\Delta m^2}{4E}\cos2\theta +
        \frac{(b_e - c_{ee})E}{2} & \frac{\Delta m^2}{4E}\sin2\theta \\
        \frac{\Delta m^2}{4E}\sin2\theta & \frac{\Delta m^2}{4E}\cos2\theta +
        \frac{(b_{\mu} - c_{\mu \mu})E}{2}
        \end{pmatrix}
    \ee
proceeds along similar lines. First of all we define an effective mixing angle
$\theta_{\mathscr{C}\text{-even}}$ in order to diagonalize the system and find a relation between the
effective mixing angle and the vacuum mixing angle
    \be
        \tan2\theta_{\mathscr{C}\text{-even}} = \frac{\Delta m^2 \sin2\theta}{(b_{\mu} - b_e +
        c_{ee}-c_{\mu\mu})E^2+\Delta m^2 \cos2\theta} \label{thetaeven},
    \ee
which reveals resonant mixing for states with an energy
    \be
        E^{\mathscr{C}\text{-even}}_{\text{res}} = \sqrt{\frac{\Delta m^2 \cos2\theta}{b_e - b_{\mu} +
        c_{\mu\mu}- c_{ee}}} \label{Ereseven}.
    \ee
Again the existence of the resonance depends on the realization of the condition $b_e - b_{\mu} + c_{\mu\mu}- c_{ee} > 0$.
The effective mass eigenvalues are then given by
    \be
        m_{3}^2 &=& \frac{\Sigma m^2}{2}-\frac{1}{2}[-(b_e+b_{\mu}-c_{ee}-c_{\mu\mu})E^2 +\kappa_{\mathscr{C}\text{-even}}], \label{mass3}\\
        m_{4}^2 &=& \frac{\Sigma m^2}{2}-\frac{1}{2}[-(b_e+b_{\mu}-c_{ee}-c_{\mu\mu})E^2 -\kappa_{\mathscr{C}\text{-even}}] \label{mass4},
    \ee
where
    \be
        \kappa_{\mathscr{C}\text{-even}}^2 = (b_e - b_{\mu}-c_{ee}+c_{\mu\mu})^2 E^4 +
        2\Delta m^2 (b_{\mu}-b_e+c_{ee}-c_{\mu\mu})\cos2\theta E^2 + (\Delta m^2)^2 \label{kappaeven}.
    \ee
The diagonalization of $h_{\text{eff}}^{\mathscr{C}\text{-even}}$ connects $\mathscr{C}$-eigenstates with mass eigenstates
    \be
        \left(\begin{array}{c} \nu_e^+ \\ \nu_{\mu}^+ \end{array}\right) = \begin{pmatrix}
        \cos\theta_{\mathscr{C}\text{-even}} & \sin\theta_{\mathscr{C}\text{-even}} \\
        -\sin\theta_{\mathscr{C}\text{-even}} & \cos\theta_{\mathscr{C}\text{-even}} \end{pmatrix}
        \left(\begin{array}{c} \nu_3 \\ \nu_4 \end{array}\right).
    \ee
In a neutrino-antineutrino oscillation model with two generations we have as many as six effective mass-squared differences and six rotation angles. Note, however, that only three of these effective mass-squared differences
are linearly independent.
\begin{figure}
\centering
\raisebox{6.5cm}{$\sin^22\theta_{\text{eff}}$}
\includegraphics[scale=1.3]{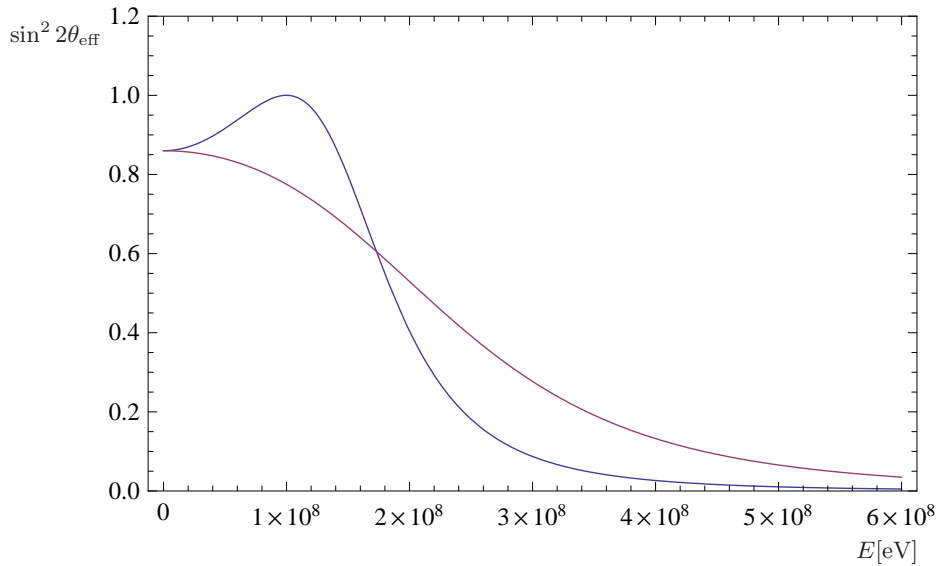}
\\{\hspace{11cm}$E[\text{eV}]$}
\caption{As in Fig. \ref{sin1}, but for $b_e=1\times 10^{-21}$, $b_{\mu}=3\times 10^{-21}$, $c_{\mu\mu}=3\times 10^{-21}$, $c_{ee}=2\times10^{-21}$ for illustrative purposes. We take $\Delta m^2 = 8 \times 10^{-5} \text{eV}^2$ as well as $\sin^22\theta = 0.86$. In this case the $\mathscr{C}\text{-odd}$-resonance exists, whereas the $\mathscr{C}\text{-even}$-resonance vanishes.} \label{sin2}
\end{figure}
Four of these effective mixing angles are absorbed into the matrix $U$ that links charge conjugation
and mass eigenstates; the other two angles are given by $\theta_{\mathscr{C}\text{-odd}}$ and $\theta_{\mathscr{C}\text{-even}}$.
\par\noindent
\newline
In addition to the fact that only states with definite $\mathscr{C}$-parity of the same kind mix, we
realize that once the parameters $b_e$, $b_{\mu}$, $c_{ee}$, $c_{\mu\mu}$ are fixed we encounter resonant
mixing for $\mathscr{C}$-odd states and $\mathscr{C}$-even states. The two conditions to be satisfied for
resonant mixing are
    \be
        b_e - b_{\mu} &<& ~~~~c_{\mu\mu} - c_{ee} \qquad \mathscr{C}\text{-odd resonances}, \label{pspace1} \\
        b_e - b_{\mu} &>& -(c_{\mu\mu} - c_{ee}) \qquad \mathscr{C}\text{-even resonances}
        \label{pspace2},
    \ee
and the resonance structure of oscillation phenomena depends on the choice of the parameters.
\par
On the one hand we may assume that
    \be
        \text{sgn} ~ (c_{\mu\mu} - c_{ee}) = +1,
    \ee
which divides the parameter space into three different regions:
For $b_e -b_{\mu}$ larger than $|c_{\mu\mu} - c_{ee}|$ only the $\mathscr{C}$-even resonance exists,
the $\mathscr{C}$-odd resonance does not exist. For $b_e -b_{\mu}$ larger than $-|c_{\mu\mu} - c_{ee}|$,
but smaller than $|c_{\mu\mu} - c_{ee}|$ the resonance conditions for both $\mathscr{C}$-even and
$\mathscr{C}$-odd states are met at the same time; oscillations reveal two resonances with distinct
resonance energies. The region of the parameter space for which $b_e - b_{\mu}$ is smaller than
$-|c_{\mu\mu} - c_{ee}|$ allows for a resonance in the $\mathscr{C}$-odd, but not in the
$\mathscr{C}$-even sector.
\par
On the other hand we might as well assume that
    \be
        \text{sgn}~(c_{\mu\mu} - c_{ee}) = -1,
    \ee
which also divides the parameter space into three regions:
For $b_e -b_{\mu}$ larger than $|c_{\mu\mu} - c_{ee}|$ we only encounter the $\mathscr{C}$-even resonance,
whereas the $\mathscr{C}$-odd resonance ceases to exist. For $b_e -b_{\mu}$ smaller than $|c_{\mu\mu} - c_{ee}|$
and larger than $-|c_{\mu\mu} - c_{ee}|$ none of the appropriate resonance conditions Eqs.~(\ref{pspace1}-\ref{pspace2}) can be met;
for this case there are no resonances. For $b_e -b_{\mu}$ smaller than
$-|c_{\mu\mu} - c_{ee}|$ the $\mathscr{C}$-odd resonance exists; the $\mathscr{C}$-even resonance does not exist.
\par\noindent
\newline
The existence of a resonance naturally divides the energy regime into three regions. We exemplify the way
to pin down the $\mathscr{C}$-flavor content of the mass branches for energies above the resonance, $E \gg E_{\text{res}}$,
at the resonance, $E = E_{\text{res}}$, and below the resonance, $E \ll E_{\text{res}}$, by examining the
relation between the effective and the standard mixing angle for $\mathscr{C}$-odd oscillations
    \be
        \tan2\theta_{\mathscr{C}\text{-odd}}(E) = \frac{\Delta m^2 \sin2\theta}{(b_e-b_{\mu}+
        c_{ee}-c_{\mu\mu})E^2+\Delta m^2 \cos2\theta}.
    \ee
For high energies we find
    \be
        \tan2\theta_{\mathscr{C}\text{-odd}}(E \to \infty) = 0,
    \ee
which means that the effective mixing angle for energies well above the resonance goes to zero.
At the resonance energy we have maximal $\mathscr{C}$-flavor mixing in a way that
    \be
        \tan2\theta_{\mathscr{C}\text{-odd}}(E = E^{\mathscr{C}\text{-odd}}_{\text{res}}) \to \infty,
    \ee
and eventually for energies well below the resonance energy we find
    \be
        \tan2\theta_{\mathscr{C}\text{-odd}}(E  \to 0) = \tan2\theta,
    \ee
i.e. we recover standard vacuum mixing. This treatment of neutrino mixing in the different energy regimes is
reminiscent to the well-known treatment of matter effect in neutrino physics \cite{Wolfenstein:1977ue, Mikheev:1986wj, Barger:1980tf}.
The case for $\mathscr{C}$-even oscillations may be treated on an equal footing.
\par\noindent
\newline
Furthermore we define the sines of the effective mixing angles via
    \be
        \sin2\theta_{\mathscr{C}\text{-odd}} = \frac{\Delta m^2\sin2\theta}{\kappa_{\mathscr{C}\text{-odd}}} \label{sinodd}
    \ee
as well as
    \be
       \sin2\theta_{\mathscr{C}\text{-even}} = \frac{\Delta m^2\sin2\theta}{\kappa_{\mathscr{C}\text{-even}}} \label{sineven}.
    \ee
The resonance energy for these expressions is easily calculated by differentiating with respect to the energy
    \be
        \left.\frac{d\sin2\theta_{\text{eff}}}{d E}\right|_{E=E_{\text{res}}}   = 0
    \ee
 and it is given by Eqs.~(\ref{Eresodd},
\ref{Ereseven}) respectively. In Fig. \ref{sin1} and Fig. \ref{sin2} the energy dependence of the effective sines is shown for different values for the CPT-violating parameters. The extremae of the shown curves correspond to the resonant energy encountered in resonant mixing between charge conjugation eigenstates. One readily infers that the resonance structures vanish for certain
choices of the parameters.
\par
\begin{figure}
\centering
\raisebox{6.0cm}{$\left|\frac{\Delta E(\text{FW$f$M})|_{b_{\mu}, b_e \neq 0}}
{\Delta E(\text{FW$f$M})|_{b_{\mu}, b_e = 0}}\right|$}
\includegraphics[scale=1.2]{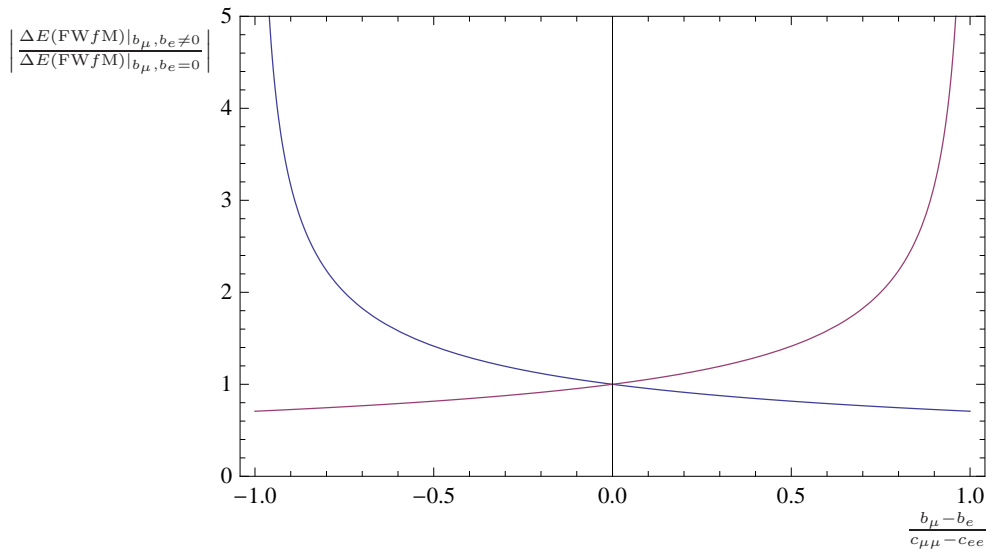}
\\{\hspace{12cm}$\frac{b_{\mu}-b_e}{c_{\mu\mu}-c_{ee}}$}
\caption{The ratio of the width of the resonances triggered by neutrino-antineutrino
oscillations and altered dispersion relations in vacuo ($b_e = b_{\mu} = 0$). The red curve shows the behavior for a $\mathscr{C}\text{-even}$ resonance; the blue curve for a $\mathscr{C}\text{-odd}$ resonance. Depending on the explicit values for the CPT-violating parameters at least one of the neutrino-antineutrino resonances reveals a narrower resonance width as compared to neutrino-neutrino oscillations with altered dispersion relations in vacuo.} \label{width}
\end{figure}
It is convenient to define the ratio of the effective mixing angles
    \be
        R(E) = \frac{\sin^22\theta_{\text{eff}}}{\sin^22\theta}
    \ee
and calculate the impact of neutrino-antineutrino mixing on altered dispersion relations in the neutrino-neutrino
mixing sector. The difference in energies at Full Width in energy at a fraction $f$ of the Maximum (FW$f$M) is
calculated to be
    \be
        \left|\frac{\Delta E(\text{FW$f$M})}{E_{\text{res}}}\right| = \left|\left[1 + \sqrt{\frac{1-f}{f}}
        \tan2\theta \right]^{\frac{1}{2}} - \left[1 - \sqrt{\frac{1-f}{f}}\tan2\theta \right]^{\frac{1}{2}}\right|
        \label{delE},
    \ee
where $\theta_{\text{eff}}$, $E_{\text{res}}$ stand either for the $\mathscr{C}\text{-odd}$ or the $\mathscr{C}\text{-even}$ resonance.
The fraction $f$ can be equivalently rewritten in terms of the ratio $R(E)$ and its resonance value $R(E_{\text{res}})$ as follows
    \be
        f = \frac{R(E)}{R(E_{\text{res}})}.
    \ee
Note that the right hand side of Eq.~(\ref{delE}) is composed of the difference of two roots which represent
the fractions of energy values on either side of the resonance peak to the value of the resonance energy. One can deduce by looking at Fig. \ref{sin1} and Fig. \ref{sin2} that both of these roots are real only above a critical value $f_{\text{crit}}$ which depends on the vacuum mixing angle $\theta$. The same conclusion can be deduced from Eq.~(\ref{delE}). The second term on the right hand side can become complex for certain choices of $f$ and therefore invalidate the notion of the FW$f$M for this particular $f$. Imposing that the expression under the square root be positive we find that $f$ should obey
    \be
        f > f_{\text{crit}} = \sin^22\theta .
    \ee
Given a solar mixing angle of $\sin^22\theta = 0.86$ this gives a critical fraction of
    \be
        f_{\text{crit}} = \frac{43}{50}.
    \ee
This in turn prohibits a sensible definition of the commonly used Full Width
at Half Maximum (FWHM) in our case for example.
\par
The ratio of the FW$f$M $\Delta E(\text{FW$f$M})|_{b_{\mu}, b_e = 0}$ for neutrino-neutrino mixing only, i.e. $b_{\mu} = b_e = 0$, and
the FW$f$M $\Delta E(\text{FW$f$M})|_{b_{\mu}, b_e \neq 0}$, i.e. $b_{\mu} \neq b_e \neq 0$, for neutrino-antineutrino mixing is given by
    \be
        \left|\frac{\Delta E(\text{FW$f$M})|_{b_{\mu}, b_e \neq 0}}{\Delta E(\text{FW$f$M})|_{b_{\mu}, b_e = 0}}\right| &=&
        \left|\frac{E_{\text{res}}|_{b_{\mu}, b_e \neq 0}}{E_{\text{res}}|_{b_{\mu}, b_e = 0}}\right| \label{ratioFWfMa} \\
        &=& \frac{1}{\sqrt{1 \pm \frac{b_{\mu}-b_e}{c_{\mu\mu}-c_{ee}}}} \label{ratioFWfMb}.
    \ee
The positive sign applies for the $\mathscr{C}\text{-odd}$ resonance; the negative sign holds for the $\mathscr{C}\text{-even}$ resonance.
The question whether the neutrino-antineutrino mixing broadens or narrows the resonance width as compared to altered dispersion relations
in vacuo, e.g. scenarios with extra-dimensional shortcuts
~\footnote
{
It can be seen by comparing the effective Hamiltonian in Eq.~(\ref{heff}) with the similar expressions from references \cite{Pas:2005rb,Hollenberg:2009bq} that the terms containing the $c_{ee}$ and $c_{\mu\mu}$ coefficients have the same energy dependence as the ones which generated the altered dispersion relations. The difference in the two is that $c_{ee}$ and $c_{\mu\mu}$ also encapsulate a directional dependence.
}
, depends on the fraction $\frac{b_{\mu}-b_e}{c_{\mu\mu}-c_{ee}}$. In Fig. \ref{width}
we show the dependence of the ratio defined in Eq.~(\ref{ratioFWfMa}) as a function of this fraction. Imposing that
the expression under the square root in Eq.~(\ref{ratioFWfMb}) be positive we recover the relation between $b_e$,
$b_{\mu}$, $c_{ee}$ and $c_{\mu\mu}$ necessary for the existence of a resonant energy. It is obvious from Fig.
\ref{width} that for certain values of the CPT-violating parameters neutrino-antineutrino oscillations lead to a
narrower resonance width as compared to neutrino-neutrino oscillations with altered dispersion relations in vacuo.
Moreover, for a fixed value of the fraction $\frac{b_{\mu}-b_e}{c_{\mu\mu}-c_{ee}}$ one of the resonances has a smaller FW$f$M value,
whereas the other resonance has a larger FW$f$M as compared to neutrino-neutrino oscillations with altered dispersion relations in vacuo.

\section{Reference frames}\label{discussion}

In order to compare different experiments, which work in different reference frames, in Lorentz-violating
models it is convenient to introduce a standard set of frames. The sun-centered celestial  equatorial frame
is conventionally used to report experimental findings \cite{Barger:2007dc} and the coefficients measured in
different experiments are related by observer Lorentz transformations. In our model it is obvious from the choice
of the non-zero CPT-violating parameters that the $Z$ direction plays a special role. We infer from the angle dependence of
    \be
        c_{ab} = 2 (c_{\text{L}})_{ab}^{TT}[1+\cos^2\Theta]
    \ee
that the only modification of $c$-type coefficients generated by direction dependence amounts to an
enhancement by a factor of at most two. For $\Theta$ equal to zero, i.e. neutrinos propagating along the $Z$ direction,
the term in square brackets is equal to two; for $\Theta$ equal to $\frac{\pi}{2}$ this factor is equal
to one.
\par
The $b$-type coefficients show a different direction dependence. The most remarkable feature in this context
turns out to be the fact that for a neutrino propagation along the $Z$ axis, i.e. $\Theta = 0$ or $\Theta = \pi$,
neutrino-antineutrino mixing is completely absent according to
    \be
        b_a = 2 \sin\Theta~g_a.
    \ee
Note here that the choice of a sun-centered coordinate frame introduces different time scales on which the angles $\Theta$
and $\Phi$ vary. These variations, however, are typically of the order of magnitude $\mathcal{O}(10^3\text{s})$, i.e. the angles typically vary
significantly within an hour or a day at least. The natural time scale of the neutrino propagation in short baseline experiments
like LSND or MiniBooNE is given by $t = L/c$ with $L$ the length of the baseline and $c$ the speed of light. Choosing $L = 10^3 \text{m}$ as a typical baseline length in short baseline experiments gives a propagation time of the neutrino between creation and detection of $t \simeq 10^{-6}\text{s}$.
We can therefore assume that the additional time dependence introduced by the motion of the earth around the sun as well as the rotation of
the earth does not influence the dynamics of neutrino-antineutrino oscillations in short baseline experiments and can hence be neglected
when it comes to the diagonalization of the effective Hamiltonian.
\par
However, the propagation time of solar neutrinos ($\sim\mathcal{O}(10^2\text{s})$) is of same order of magnitude with the typical time scales of variation of the celestial colatitude and longitude. As a result a time dependence through $\Theta$ and $\Phi$ is induced in the effective Hamiltonian from Eq. (\ref{heff}). In order to do a similar analysis for solar neutrinos one has to solve the Schr\"odinger equation with a time dependent Hamiltonian which would be more involved than the diagonalization procedure considered in this work.
\par
Another issue in this context is the time dependence of $\Theta$ and $\Phi$ during the total running time of the experiments.
Within this model correlations between resonant enhancements in oscillations and the orientation of a certain
experiment in the sun-centered celestial equatorial frame are present.
A detailed analysis of this effect lies beyond the scope of this letter though and is left for future work.

\section{From charge conjugation to flavor eigenstates}\label{discussion2}

The translation of $\mathscr{C}$-eigenstates into common flavor eigenstates
measured in neutrino oscillation experiments is another important issue. We can construct the matrix $V$ that diagonalizes the effective
Hamiltonian $h_{\text{eff}}$ and translates the common flavor eigenstates $\nu_e,~\nu_{\mu},~\nu_e^c,~\nu_{\mu}^c$ into the
mass eigenbasis $\nu_1,~\nu_2,~\nu_3,~\nu_4$ by
    \be
        V &=& \frac{1}{\sqrt{2}} \begin{pmatrix} 1 & 0 & 1 & 0 \\ 0 & 1 & 0 &1 \\ -1 & 0 & 1 & 0 \\
        0 & -1 & 0 & 1 \end{pmatrix}
        \begin{pmatrix} \cos\theta_{\mathscr{C}\text{-odd}} & \sin\theta_{\mathscr{C}\text{-odd}}
        & 0 & 0 \\ -\sin\theta_{\mathscr{C}\text{-odd}} & \cos\theta_{\mathscr{C}\text{-odd}} & 0 & 0
        \\ 0 & 0 & \cos\theta_{\mathscr{C}\text{-even}} & \sin\theta_{\mathscr{C}\text{-even}} \\
        0 & 0 & -\sin\theta_{\mathscr{C}\text{-even}} & \cos\theta_{\mathscr{C}\text{-even}} \end{pmatrix}
        \nonumber \label{V} \\
        &=& \frac{1}{\sqrt{2}} \begin{pmatrix} \cos\theta_{\mathscr{C}\text{-odd}} & \sin\theta_{\mathscr{C}\text{-odd}} &
        \cos\theta_{\mathscr{C}\text{-even}} & \sin\theta_{\mathscr{C}\text{-even}} \\ -\sin\theta_{\mathscr{C}\text{-odd}} &
        \cos\theta_{\mathscr{C}\text{-odd}} & -\sin\theta_{\mathscr{C}\text{-even}} & \cos\theta_{\mathscr{C}\text{-even}}
        \\ -\cos\theta_{\mathscr{C}\text{-odd}} & -\sin\theta_{\mathscr{C}\text{-odd}} & \cos\theta_{\mathscr{C}\text{-even}} &
        \sin\theta_{\mathscr{C}\text{-even}} \\
        \sin\theta_{\mathscr{C}\text{-odd}} & -\cos\theta_{\mathscr{C}\text{-odd}} & -\sin\theta_{\mathscr{C}\text{-even}} &
        \cos\theta_{\mathscr{C}\text{-even}} \end{pmatrix}.
    \ee
This amounts to a translation between flavor and mass eigenstates given by
    \be
        \left(\begin{array}{c} \nu_e \\ \nu_{\mu} \\ \nu_e^c \\ \nu_{\mu}^c \end{array}\right) =
        \frac{1}{\sqrt{2}} \begin{pmatrix} \cos\theta_{\mathscr{C}\text{-odd}} & \sin\theta_{\mathscr{C}\text{-odd}} &
        \cos\theta_{\mathscr{C}\text{-even}} & \sin\theta_{\mathscr{C}\text{-even}} \\ -\sin\theta_{\mathscr{C}\text{-odd}} &
        \cos\theta_{\mathscr{C}\text{-odd}} & -\sin\theta_{\mathscr{C}\text{-even}} & \cos\theta_{\mathscr{C}\text{-even}}
        \\ -\cos\theta_{\mathscr{C}\text{-odd}} & -\sin\theta_{\mathscr{C}\text{-odd}} & \cos\theta_{\mathscr{C}\text{-even}} &
        \sin\theta_{\mathscr{C}\text{-even}} \\
        \sin\theta_{\mathscr{C}\text{-odd}} & -\cos\theta_{\mathscr{C}\text{-odd}} & -\sin\theta_{\mathscr{C}\text{-even}} &
        \cos\theta_{\mathscr{C}\text{-even}} \end{pmatrix}
        \left(\begin{array}{c} \nu_1 \\ \nu_2 \\ \nu_3 \\ \nu_4 \end{array}\right),
    \ee
whereas the diagonal Hamiltonian $h_{\text{eff}}^{\text{diag}}$ is obtained via
    \be
        h_{\text{eff}}^{\text{diag}} = V^{\dagger} h_{\text{eff}} V.
    \ee
\par\noindent
\newline
The probability of oscillation is thus given by squaring the amplitude of oscillation
    \be
        P(\beta \to \alpha) &=& \left|\sum_i V_{\beta i} e^{-iE_it} \left(V^{\dagger}\right)_{i\alpha}\right|^2 \\
        &=& \sum\limits_{ij} V_{\beta i} V_{\alpha j} \left(V^{\dagger}\right)_{i \alpha}
            \left(V^{\dagger}\right)_{j\beta} e^{-i(E_i-E_j)t},\label{oscprob}
    \ee
where $E_i$ are the effective energy eigenvalues of the associated Hamiltonian $h_{\text{eff}}$; $\alpha$ and $\beta$ stand for the four different neutrino species involved, i.e. $\alpha,~\beta = \nu_e,~\nu_{\mu},~\nu_e^c,~\nu_{\mu}^c$.
\par\noindent
In the limit in which
the CPT-violating coefficients are absent one would not expect any residual impact of $U$ on the oscillation probability, since the Hamiltonian is already block-diagonal. Neutrino-antineutrino oscillations vanish in this limit. We will now show that this is indeed the case:
\par
Without CPT-violating terms, the Hamiltonian $h_{\text{eff}}$ is already block-diagonal and the following two relations
    \be
        \left[U, R\right] &=& 0, \label{com1} \\
        \left[U, h_{\text{eff}}\right] &=& 0 \quad \Leftrightarrow \quad \sum\limits_{jk}U_{ij} h^{\text{eff}}_{jk} \left(U^{\dagger}\right)_{kl} = \sum\limits_{jk}h^{\text{eff}}_{jk} \delta_{ji}\delta_{kl} \label{com2}
    \ee
hold. Using the matrix $U$ as given in Eq. (\ref{U}) as well as the rotation matrix
    \be
        R = \begin{pmatrix} \cos\theta_{\mathscr{C}\text{-odd}} & \sin\theta_{\mathscr{C}\text{-odd}}
        & 0 & 0 \\ -\sin\theta_{\mathscr{C}\text{-odd}} & \cos\theta_{\mathscr{C}\text{-odd}} & 0 & 0
        \\ 0 & 0 & \cos\theta_{\mathscr{C}\text{-even}} & \sin\theta_{\mathscr{C}\text{-even}} \\
        0 & 0 & -\sin\theta_{\mathscr{C}\text{-even}} & \cos\theta_{\mathscr{C}\text{-even}}\end{pmatrix},
    \ee
we write $V$ as follows
    \be
        V = UR \label{Vs}.
    \ee
Furthermore it can easily be seen that for $b_e = b_{\mu} = 0$, the diagonalization of the Hamiltonian only introduces
one effective mixing angle, i.e. $\theta_{\mathscr{C}\text{-odd}} = \theta_{\mathscr{C}\text{-even}} \equiv \theta
_{\text{eff}}$, since the two $2\times 2$ blocks in the upper left and lower right corner of the full $4 \times 4$
effective Hamiltonian are identical.
Making use of Eqs. (\ref{com1}, \ref{com2}, \ref{Vs}) we find
    \be
        P(\beta \to \alpha) &=& \sum\limits_{ij} \left(RU\right)_{\beta i} \left(RU\right)_{\alpha j} \left(U^{\dagger}R^{\dagger}\right)_{i \alpha}
    \left(U^{\dagger}R^{\dagger}\right)_{j\beta} e^{-i(E_i-E_j)t} \\
                        &=& \sum\limits_{klmn} R_{\beta k} R_{\alpha l} \left(R^{\dagger}\right)_{m\alpha}
                        \left(R^{\dagger}\right)_{n\beta} \left(\sum\limits_{i} U_{k i} e^{-iE_it} \left(U^{\dagger}\right)_{im} \right) \left(\sum\limits_{j} U_{lj}  e^{iE_jt} \left(U^{\dagger}\right)_{jn} \right) \label{theEs1}\\
                        &=& \sum\limits_{klmn} R_{\beta k} R_{\alpha l} \left(R^{\dagger}\right)_{m\alpha}
                        \left(R^{\dagger}\right)_{n\beta} e^{-iE_kt} \delta_{km} e^{iE_lt} \delta_{ln} \label{theEs2}\\
                        &=& \sum\limits_{kl} R_{\beta k} R_{\alpha l} \left(R^{\dagger}\right)_{k\alpha}
                        \left(R^{\dagger}\right)_{l\beta} e^{-i(E_k-E_l)t}.
    \ee
In order to get from Eq. (\ref{theEs1}) to Eq. (\ref{theEs2}) the fact that $E_1$ and $E_3$ as well as
$E_2$ and $E_4$ are equal was used. This assumption is sensible -- since for CPT-conserving neutrino oscillations there is only
one effective mass-squared difference -- and readily corroborated by looking at Eqs. (\ref{mass1}, \ref{mass2}, \ref{mass3}, \ref{mass4}).
In the absence of Lorentz-violating effects as well, when also the $C$-type coefficients are absent, we are left with standard vacuum oscillations by setting
$\theta_{\text{eff}} = \theta_{\text{vac}}$.
\par\noindent
\newline
Eventually we remark that the resonances encountered in neutrino-antineutrino mixing might be suitable to
accommodate the LSND and MiniBooNE anomalies. A careful treatment would imply a detailed analysis of the
oscillation probability though. It is important to notice that the resonance structures analyzed arise not due to the transition between flavor eigenstates, but rather in a basis of $\mathscr{C}$-eigenstates. The analysis
of the resonance widths explicitly refers to resonance structures in $\mathscr{C}$-even and $\mathscr{C}$-odd mixing only. Nevertheless, the mixing matrix $V$ can be parameterized in terms of the $\mathscr{C}\text{-even}$ and $\mathscr{C}\text{-odd}$ mixing angles.
It is therefore at least conceivable that the property of narrower resonances due to the presence of CPT violation will be retained when
switching to the flavor basis. We leave a detailed discussion of this issue and its phenomenology for future work \cite{HMP2}.
\par
An estimate of the typical order of magnitude for the Lorentz-violating parameters involved can be given as follows.
A resonance energy $E^2_{\text{res}} \gg \Delta m^2$ is sought so the
ratio
    \be
        \frac{E^2_{\text{res}}}{\Delta m^2} = \frac{\cos2\theta}{\pm (b_{\mu}-b_e)+c_{\mu\mu}-c_{ee}}
    \ee
must be large, or equivalently, $\left[\pm (b_{\mu}-b_e)+c_{\mu\mu}-c_{ee}\right]$ must be small. As a numerical
example we consider a resonance energy of $E_{\text{res}} = 200~\text{MeV}$; given the solar mass-squared difference of $\Delta m^2 = 8 \times 10^{-5}~\text{eV}^2$ this amounts to
    \be
        \left[\pm (b_{\mu}-b_e)+c_{\mu\mu}-c_{ee}\right] \sim \mathcal{O}\left(10^{-21}\right).
    \ee
Again the positive sign holds for a $\mathscr{C}$-odd resonance, whereas the negative sign applies for a
$\mathscr{C}$-even resonance.

\section{Conclusion}\label{conclusion}

We have studied neutrino-antineutrino oscillation phenomena in a two generation framework with CPT- and Lorentz-violating coefficients.
Such a model can potentially solve several major problems encountered in attempts to explain the MiniBooNE and
LSND anomalies in terms of active-sterile neutrino oscillations with altered dispersion relations:
\par
Active neutrino disappearance due to oscillations into sterile neutrinos does not arise since the model is not based on the existence of a sterile neutrino state. In addition CPT violation distinguishes particles and antiparticles such that resonance peaks for neutrinos and antineutrinos are not necessarily identical as it can be inferred from the probability of oscillation in Eq. (\ref{oscprob}). A recent analysis of experimental neutrino data \cite{Karagiorgi:2009nb} in $(3+1)$ and $(3+2)$ neutrino scenarios also reveals incompatibilities between
neutrino and antineutrino data sets which in turn might favor alternative approaches like CPT violation to reconcile the MiniBooNE and LSND anomalies. The hint that the signal observed at MiniBooNE looks more like a
$\nu_{\mu} \to \bar{\nu}_e$ conversion than $\nu_{\mu} \to \nu_e$ events can be accommodated via neutrino-antineutrino oscillations.
\par
It has been found that a conveniently simple choice of non-zero CPT-violating coefficients provides a workable model which already entails interesting phenomenological consequences.
In particular the model for neutrino-antineutrino oscillations under consideration in a CPT-violating framework gives rise to new vacuum resonances \cite{Barger:2000iv} which are suitably described in terms of
$\mathscr{C}$-even and $\mathscr{C}$-odd states. Resonant mixing as defined occurs between $\mathscr{C}$-flavor eigenstates rather than between common flavor eigenstates.
Depending on the parameter space of the CPT-violating coefficients it is possible to have none, one or two resonances. These resonances are related to the mixing of $\mathscr{C}$-flavor eigenstates. One finds that at least one of the neutrino-antineutrino resonances reveals a narrower resonance width as compared to neutrino-neutrino oscillations with altered dispersion relations in vacuo.
Depending on the choice of parameters, the model predicts interesting daily and seasonal variations of neutrino
oscillation observables, which result from the Earth's motion with respect to a preferred frame implied by a
Lorentz-violating background field.
A detailed analysis of both the direction dependence of CPT-violating coefficients as well as the flavor oscillation probability will possibly shed light on neutrino oscillation anomalies such as LSND and MiniBooNE.

\acknowledgments
We thank Danny van Dyk for a useful discussion in the early stage of this work as well as Thomas J. Weiler
for useful remarks.

\end{document}